# Ultrafast dynamics and surface plasmon properties of Silicon upon irradiation with mid-infrared femtosecond laser pulses

E. Petrakakis [1,2], G. D. Tsibidis [1*] and E. Stratakis [1,2]

[1] *Institute of Electronic Structure and Laser (IESL), Foundation for Research and Technology (FORTH), N. Plastira 100, Vassilika Vouton, 70013, Heraklion, Crete, Greece*

[2] *Materials Science and Technology Department, University of Crete, 71003 Heraklion, Greece*

We present a theoretical investigation of the yet unexplored ultrafast processes and dynamics of the produced excited carriers upon irradiation of Silicon with femtosecond pulsed lasers in the mid-infrared (mid-IR) spectral region. The evolution of the carrier density and thermal response of the electron-hole and lattice subsystems are analysed for various wavelengths $\lambda_L$ in the range between 2.2 μm and 3.3 μm where the influence of two and three-photon absorption mechanisms is explored. The role of induced Kerr effect is highlighted and it manifests a more pronounced influence at smaller wavelengths in the mid-IR range. Elaboration on the conditions that leads to surface plasmon (SP) excitation indicate the formation of weakly bound SP waves on the material surface. The lifetime of the excited SP is shown to rise upon increasing wavelength yielding a larger than the one predicted for higher laser frequencies. Calculation of damage thresholds for various pulse durations $\tau_p$ show that they rise according to a power law ($\sim \tau_p^{\zeta(\lambda_L)}$) where the increasing rate is determined by the exponent $\zeta(\lambda_L)$. Investigation of the multi-photon absorption rates and impact ionization contribution at different $\tau_p$ manifests a lower damage for $\lambda_L$=2.5 μm compared to that for $\lambda_L$=2.2 μm for long $\tau_p$.

## I. INTRODUCTION

Over the past decades, the use of ultra-short pulsed laser sources for material processing and associated laser driven physical phenomena have received considerable attention due to the important technological applications, in particular in industry and medicine [1-10]. Various types of surface structures generated by laser pulses and more specifically, the so-called laser-induced periodic surface structures (LIPSS) on solids have been studied extensively [11-24]. Previous theoretical approaches or experimental observations related to the understanding of the underlying physical mechanisms for the formation of these structures were performed in a variety of conditions. To explain the physical origin of LIPSS formation, it is important to note that following irradiation with ultrashort pulses, a series of multiphysical phenomena take place (i.e. energy absorption, electron excitation and relaxation processes, phase transition/mechanical effects and solidification) [25-31]. With respect to the formation of subwavelegth structures of size comparable to the laser wavelength $\lambda_L$, various mechanisms have been proposed to account for the production of periodic structures: interference of the incident wave with an induced scattered wave [12, 14, 17], or with a surface plasmon wave (SP) [13, 16, 32-37], or due to self-organisation mechanisms [38].

A key characteristic of the aforementioned surface modification mechanisms and the associated physical processes is that they were explored for laser pulses in a region between the visible and near-infrared spectrum ($\lambda_L <$ 1.5 μm). Nevertheless, to the best of our knowledge, laser machining in the mid-IR ($\lambda_L > 2$ μm) is a yet unexplored field [39, 40]; the motivation to investigate the response of the irradiated semiconductor and relevant surface effects in the mid-IR region originates from the challenging opportunities in photonics for mid-IR radiation [41-43]. A fundamental question is whether the underlying physics that characterizes laser-matter interaction is similar for mid-IR and lower spectral regions. One characteristic example is that the proposed physical mechanism for the formation of sub-wavelength structures (i.e. surface plasmon excitation) appears to behave differently if mid-IR sources are used [41]; for instance, the response of plasmons in the visible is largely dominated by (ohmic) losses in metals and the reaction time of the electrons. The former is closely related to the fact that the optical field of surface plasmons is weakly bound at mid-IR irradiation unlike the confinement at visible wavelengths with direct implications in the energy absorption and potential applications [44, 45]. Therefore, from both a fundamental and application point of view, it is important to explore the following physical processes for irradiation in the mid-IR regime: (i) energy absorption, SP-excitation, and confinement of the SP-fields to the surface, (ii) morphological effects following SP-excitation (i.e. interference of the weakly confined SP with the electric field of the incident beam), (iii) damage onset and relevant threshold (i.e. corresponding to the fluence that leads to the onset of surface modification.

To account for the influence of mid-IR photons on the physical processes taking place upon irradiation of a semiconductor, a number of critical factors should be considered: (i) the transparency of the material at larger $\lambda_L$ (Silicon is expected to behave as a 'wide band-gap' semiconductor) [41] (ii) the significance of nonlinear processes such as Kerr effect or multi-photon absorption [42, 43, 46], (iii) the role of the wavelength in the modulation of the optical parameters, and (iv) the spatial and temporal width of excited Surface Plasmon field in mid-IR. Furthermore, one characteristic condition related to the 'metallisation' of the irradiated material is the vanishing of the real part of the dielectric constant $\varepsilon$ of the semiconductor for a critical value of the carrier density $N_e^{(cr)}$ [47]. The above condition leads to a value for $N_e^{(cr)}$ that is inversely proportional to $\lambda_L^2$ [48]. In this context, assuming that SP excitation and coupling with the incident beam is the predominant mechanism for the LIPSS formation [16, 37] which, in turn, requires that $Re(\varepsilon)<-1$, an interesting question that needs to be explored is whether SP-excitation and LIPSS formation can be performed at substantially lower free carrier densities or laser beam energies.

Therefore, to provide a detailed investigation of the response of the material upon irradiation with mid-IR pulses, a parametric analysis of the influence of the laser parameters (i.e. fluence, pulse duration, wavelength) on the optical parameters of the material, carrier dynamics, thermal effects and the induced morphological changes is required. To this end, we present an extension of the well-established theoretical model that describes ultrafast dynamics in semiconductors, to account in this case for Silicon (Si), for excitation and electron-phonon relaxation upon irradiation with ultrashort pulsed lasers in mid-IR in the range 2.2 μm ≤ $\lambda_L$ ≤ 3.3 μm (Section II). The theoretical framework is coupled to a module that accounts for the formation of SP excitation by predicting the laser conditions for the production of sufficiently high density of excited carriers. Section III explains the details of the numerical algorithm used in this work. A detailed analysis of the results the theoretical model yields is presented in Section IV by estimating the optical parameter variation, damage thresholds and SP wave periodicities in various laser conditions. Concluding remarks follow in Section V.

## II. THEORETICAL MODEL

### a. Energy and Particle Balance equations

Following irradiation of Si with mid-IR femtosecond pulses in the range 2.2 μm ≤ $\lambda_L$ ≤ 3.3 μm, it is assumed that two-photon and/or three-photon absorption mechanisms are sufficient to excite carriers from the valence to the conduction band while higher order photon processes are less likely to occur. On the other hand, (linear) free carrier photon absorption can increase the electron energy (but not the number of the excited carriers) while Auger recombination and impact ionization processes lead to decrease or increase of the carriers in the conduction band, respectively.

To describe the carrier excitation and relaxation processes, the relaxation time approximation to Boltzmann's transport equation [16, 25, 37, 49-51] is employed to determine the spatial ($\vec{r}=(x,y,z)$) and temporal dependence ($t$) of the carrier density number, carrier energy and lattice energy; based on this picture, the following set of coupled (nonlinear) energy and particle balance equations are used to derived the evolution of the carrier density number $N_e$, carrier temperature $T_c$ and lattice temperature $T_L$

$$C_c \frac{\partial T_c}{\partial t} = -\frac{C_c}{\tau_e}(T_c - T_L) + S(\vec{r},t)$$

$$C_L \frac{\partial T_L}{\partial t} = \vec{\nabla} \cdot (K_L \vec{\nabla} T_L) + \frac{C_c}{\tau_e}(T_c - T_L) \quad (1)$$

$$\frac{\partial N_e}{\partial t} = \frac{\beta_{TPA}}{2\hbar\omega_L} I^2(\vec{r},t) + \frac{\gamma_{TPA}}{3\hbar\omega_L} I^3(\vec{r},t) - \gamma N_e^3 + \theta N_e - \vec{\nabla} \cdot \vec{J}$$

where $C_c$ ($C_L$) is the carrier (lattice) heat capacity, $k_e$ ($k_h$) is the heat conductivities of the electron (holes), $\hbar\omega_L$ stands for the photon energy, $\beta_{TPA}$ and $\gamma_{TPA}$ correspond to the two- and three-photon absorption coefficients, respectively, $\gamma$ is the coefficient for Auger recombination, $\theta$ is the impact ionization coefficient, and $\tau_e$ is the carrier-phonon energy relaxation time. Other quantities appear in Eq.1 are the carrier current density $\vec{J}$, the heat current density $\vec{W}$ and $S$ provided by the following expressions

$$S(\vec{r},t) = \alpha_{FCA} I(\vec{r},t) + \beta_{TPA} I^2(\vec{r},t) + \gamma_{TPA} I^3(\vec{r},t) - \vec{\nabla} \cdot \vec{W}$$
$$- \frac{\partial N_e}{\partial t}(E_g + 3k_B T_e) - N_e \left( \frac{\partial E_g}{\partial T_L} \frac{\partial T_L}{\partial t} + \frac{\partial E_g}{\partial N_e} \frac{\partial N_e}{\partial t} \right)$$
$$\vec{W} = (E_g + 4k_B T_e)\vec{J} - (k_e + k_h)\vec{\nabla} T_e \quad (2)$$
$$\vec{J} = -D\left( \vec{\nabla} N_e + \frac{N_e}{2k_B T_e}\vec{\nabla} E_g + \frac{N_e}{2T_e}\vec{\nabla} T_e \right)$$

where $\alpha_{FCA}$ is the free carrier absorption coefficient, $D$ stands for the ambipolar carrier diffusivity, $k_B$ stands for the Boltzmann constant. Values of all parameters and coefficients used in this work are presented in Ref. [16, 50] and Table I. In previous studies, simulations manifested that although heat dissipation and particle transport are expected to lower the damage threshold predictions (results were given for $\lambda_L$ = 800 nm and $\tau_p$ < 1 ps [25, 49, 50]), ignoring these effects does not produce substantial changes to the material response. Nevertheless, in the current work, to perform a rigorous approach, the complete model was used despite it is computationally more demanding.

The energy flux $I(\vec{r},t)$ at a given thickness $z$ inside the target (Eqs.(1-2)) is obtained by considering the laser energy propagation loss due to two-, three- photon and free carrier absorption, respectively [25]

$$\frac{\partial I(\vec{r},t)}{\partial z} = -\alpha_{FCA} I(\vec{r},t) - \beta_{TPA} I^2(\vec{r},t) - \gamma_{TPA} I^3(\vec{r},t) \quad (3)$$

assuming that the laser beam is Gaussian both temporally and spatially while the transmitted laser intensity at the incident surface is expressed in the following form

$$I(x,y,z=0,t) = \frac{2\sqrt{\ln 2} E_p (1-R(z=0,t))}{\sqrt{\pi}\tau_p} e^{-\left(\frac{2(x^2+y^2)}{R_0^2}\right)} e^{-4\ln 2 \left(\frac{t-t_0}{\tau_p}\right)^2} \quad (4)$$

where $E_p$ is the fluence of the laser beam and $\tau_p$ is the pulse duration (i.e. full width at half maximum), $R_0$ is the irradiation spot-radius (distance from the centre at which the intensity drops to $1/e^2$ of the maximum intensity, and $R$ is the reflectivity while irradiation under normal incidence was assumed.

**b. Optical properties of the irradiated material**

The computation of the free carrier absorption coefficient and the reflectivity are derived from the dielectric constant of the material (based on the Drude model assuming also corrections due to band and state filling [47]), $\varepsilon'$,

$$\varepsilon' = 1 + (\varepsilon_{un} - 1)\left(1 - \frac{N_e}{N_v}\right) - \frac{e_c^2 N_e}{\varepsilon_0 \omega_L^2} \frac{1}{\left(1 + i\frac{1}{\omega_L \tau_{col}}\right)} \left(\frac{1}{m^*_{e-cond}} + \frac{1}{m^*_{h-cond}}\right) \quad (5)$$

where $\varepsilon_{un}$ is the dielectric constant of the unexcited material at $\lambda_L$ (for $\lambda_L \geq 2.5$ μm [52], and for $\lambda_L \leq 2.5$ μm [53]), $e_c$ is the electron charge, $m^*_{e-cond} = 0.26\, m_{e0}$  $m^*_{h-cond} = 0.37\, m_{e0}$ are the optical effective masses of the carriers [25, 50] for conductivity calculations, $m_{e0}$ is the electron mass, $\varepsilon_0$ is the permittivity of vacuum, $N_v$ corresponds to the valence band carrier density (~$5\times10^{22}$ cm$^{-3}$) and $\tau_{col}$ stands for the carriers (electron-hole) collision time. Although, more complex expressions were used in previous reports [50] to compute $\tau_{col}$ based on the electron-phonon, hole-phonon and electron-hole collision, in this work, a constant value, $1/\tau_{col}$ ~ $1.5\times10^{14}$ s$^{-1}$ is assumed in the current simulations [16, 32, 37]. The reflectivity and free carrier absorption coefficients are given by the following expressions

$$\alpha_{FCA}(x,y,z,t) = \frac{2\omega_L k}{c}$$
$$R(x,y,z=0,t) = \frac{(1-n)^2 + k^2}{(1+n)^2 + k^2} \quad (6)$$

where $c$ stands for the speed of light while $n$ and $k$ ($n = n_0 + n_2 I$, $\varepsilon = (n+ik)^2$) are the refractive index and extinction coefficient of the material, respectively [55]

$$\varepsilon = \varepsilon' + \Delta\varepsilon_{kerr}$$
$$\Delta\varepsilon_{kerr} = 2n_0 n_2 I + (n_2 I)^2 \quad (7)$$

while $n_2$ is the Kerr coefficient that corresponds to the nonlinear part of the refractive index due to Kerr effect; it is related to the real part of the primary third-order susceptibility $\chi^{(3)'}$ through the following expression [46, 56]

$$n_2 \simeq \frac{3}{4\varepsilon_0 c (n_0(N_e))^2} \mathrm{Re}\left(\chi^{(3)'}\right) \quad (8)$$

while $n_0$ stands for the refractive index for $n_2=0$.

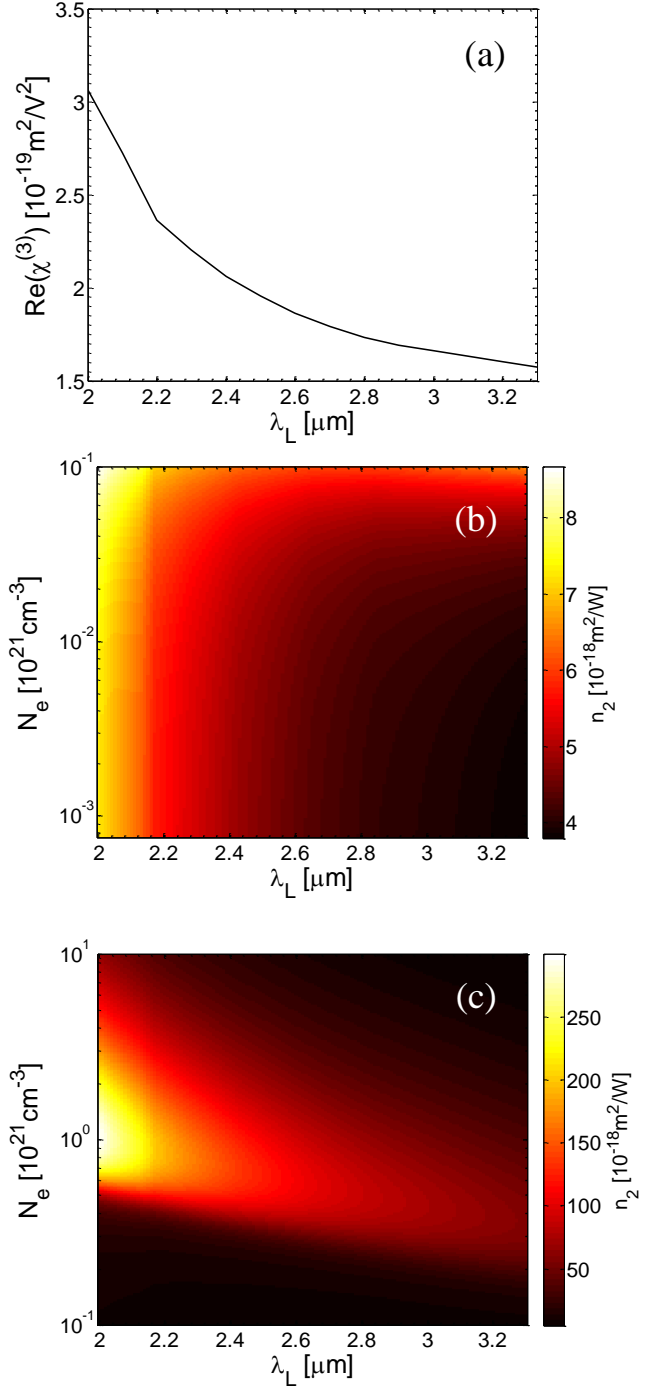

FIG. 1 (Color Online): (a) Computation of $\text{Re}(\chi^{(3)'})$ as a function of the wavelength [42, 46, 54]. Dependence of Kerr coefficient on $\lambda_L$ and $N_e$ is illustrated for low (b) and high carrier (c) densities.

The computation of $\mathrm{Re}\left(\chi^{(3)'}\right)$ is performed by following a fitting procedure on the averaged experimental data [46] in Refs.[42] and [54] (Fig.1a) while Eqs.5 and 8 indicate a carrier density dependent Kerr coefficient (Fig.1b,c). Results show significant values for the $\mathrm{Re}\left(\chi^{(3)'}\right)$ and $n_2$ within the spectral region explored in this work (2.2 μm ≤ $\lambda_L$ ≤ 3.3 μm), especially for low (Fig.1b) and high (Fig.1c) carrier density and therefore, it is important to explore the role of the nonlinear part of the refractive index in the optical and thermal response of the material.

**c. Surface Plasmon Excitation**

In previous works, it has been reported that the excitation of SP and its interference with the incident beam constitute one of the dominant mechanisms that aim to explain the formation of LSFL [16, 27, 37, 45, 57]. The dispersion relation for the excitation of SP is derived by the boundary conditions (continuity of the electric and magnetic fields at the interface between a metallic and dielectric material) ($\varepsilon_d = 1$). Therefore, a requirement for a semiconductor to obey the above relation and conditions [16, 45] is that $Re(\varepsilon) < -1$ and the computed SP wavelength $\lambda_S$ is given by

$$\lambda_S = \frac{\lambda_L}{\mathrm{Re}\sqrt{\left(\frac{\varepsilon\varepsilon_d}{\varepsilon + \varepsilon_d}\right)}} \quad (9)$$

The condition $Re(\varepsilon) < -1$ and Eq.5 can be used to derive the range of values of the excited carrier densities that lead to SP excitation. Nevertheless, special attention is required for the evaluation of the correlation of the induced SP wavelength with the carrier densities compared to the standard procedure followed for lower laser beam wavelengths (for example, at $\lambda_L = 800$ nm) [16, 37] where Kerr effect is negligible. This is due to the fact that the dielectric constant of the excited material depends both on $N_e$ and $I$ due the presence of $I$ in the nonlinear part of the refractive index. In principle, the intensity value directly influences the value of the carrier density, which implies that $I$ is an independent parameter. Hence, the evaluation of $\lambda_S$ requires the solution of Eqs.1-9.

### III. NUMERICAL SIMULATION

Due to the inherent complexity of the set of nonlinear equations presented in Section II, an analytical solution of the equations that yield the spatial/temporal dependence of the thermal and optical properties of the irradiated material is not feasible and therefore a numerical scheme is pursued. Numerical simulations have been performed using a finite difference method while the discretization of time and space has been chosen to satisfy the Neumann stability criterion. Furthermore, it is assumed that on the boundaries, von Neumann boundary conditions are satisfied and heat losses at the front and back surfaces of the material are negligible. The initial conditions are $T_e(t=0) = T_L(t=0) = 300$ K, and $N_e = 10^{12}$ cm$^{-3}$ at $t=0$. The parameters for Si used in the simulation are summarised in Table I. The (peak) fluence is $E_p\left(\equiv \sqrt{\pi}\tau_p I_0/\left(2\sqrt{ln2}\right)\right)$, where $I_0$ stands for the peak intensity. While the set of equations Eqs.1-8 are designed to provide a 3-D solution, for the sake of simplicity, and for the objectives of the present study, equations are solved across $z=0$ (where the energy deposition is higher and therefore surface damage is more pronounced). A common approach followed to solve similar problems is through the employment of a staggered grid finite difference method which is found to be effective in suppressing numerical oscillations. Temperatures ($T_e$ and $T_L$) and carrier densities ($N_e$) are computed at the centre of each element while time derivatives of the displacements and first-order spatial derivative terms are evaluated at locations midway between consecutive grid points. Similarly, the carrier current density $\vec{J}(x, y, z, t)$ and the heat current density $\vec{w}(x, y, z, t)$ are evaluated at the above locations rather than at the centres of each element. It is also assumed that $\vec{J}(z=0, t)$ and $\vec{w}(z=0, t)$ at the front ($z=0$) and back surfaces vanish.

Kerr effect is more pronounced at lower laser wavelengths (Fig.1), the range of laser frequencies for which the nonlinear part of the refractive index becomes significant should be highlighted [46]. It is evident that as the laser wavelength decreases, the impact of the Kerr effect is also expected to affect the ultrafast dynamics and thermal response. Firstly, the nonlinearity of the material's refractive index leads to a variation of the absorbed energy and the temporal evolution of the reflectivity (Fig.2a,b, for $\lambda_L = 2.2$ μm). This variation is more pronounced at smaller wavelengths (Fig.2c) while at larger wavelengths in the mid-IR region ($\leq 3.3$ μm), the Kerr effect is weaker and

## IV. RESULTS AND DISCUSSION

TABLE I. Model parameters for Si.

| Quantity | Symbol (Units) | Value |
|---|---|---|
| Electron-hole pair heat capacity [16, 50] | $C_c$ (J/m$^3$ K) | $3N_e k_B$ |
| Lattice heat capacity [16] | $C_L$ (J/m$^3$ K) | $10^6 \times (1.978 + 3.54 \times 10^{-4} T_L - 3.68 T_L^{-2})$ |
| Electron (hole) heat conductivity [16, 50] | $k_e$ ($k_h$) (W/m K) | $1.602 \times (-0.347 + 4.45 \times 10^{-3} T_e)$ |
| Lattice heat conductivity [16] | $K_L$ (W/m K) | $1585 \times 10^2 T_L^{-1.23}$ |
| Melting Temperature [47] | $T_{melt}$ (K) | 1687 |
| Band gap energy [16] | $E_g$ (J) | $1.602 \times 10^{-19} \times [1.16 - 7.02 \times 10^{-4} T_L^2 / (T_L + 1108) - 1.5 \times 10^{-8}$ (cm) $N_e^{1/3}]$ |
| Two-photon absorption [46] | $\beta_{TPA}$ (m/W) | Fitting ($\beta_{TPA}$ = 0.25 cm/GW for $\lambda_L$ = 2.2 μm, and 0 for $\lambda_L$ = 2.5 μm and $\lambda_L$ = 3.3 μm) |
| Kerr coefficient $\chi^{(3)'}$ [42, 46, 54] | $n_2$ (m$^2$/GW) | Fitting (Fig.1) |
| Three-photon absorption [58] | $\gamma_{TPA}$ (m$^3$/GW$^2$) | $1.54 \times 10^{-9}$ ($E_{ga}/\hbar\omega_L$)$^9$ ($\hbar\omega_L / E_{ga} - 1/3$)$^2$, where $E_{ga}=E_g$ at 300K ($\gamma_{TPA}$ = 0.020 cm$^3$/GW$^2$ for $\lambda_L$ = 2.2 μm, $\gamma_{TPA}$ = 0.0276 cm$^3$/GW$^2$ for $\lambda_L$ $\gamma_{TPA}$ = 0.0017 cm$^3$/GW$^2$ for $\lambda_L$ = 3.3 μm) |
| Auger recombination coefficient [16] | $\gamma$ (m$^6$/s) | $3.8 \times 10^{-43}$ |
| Ambipolar carrier diffusivity [16, 25, 50] | $D$ (m$^2$/s) | $18 \times 10^{-4} \times 300/T_L$ |
| Impact ionisation coefficient [16] | $\theta$ (s$^{-1}$) | $3.6 \times 10^{10} e^{-1.5 E_g / k_B T_c}$ |
| Carrier-phonon relaxation time [16, 37, 49, 59] | $\tau_e$ (s) | $\tau_{e0} \left[1 + \left(\frac{N_e}{N_{th}}\right)^2\right]$, $\tau_{e0}$ =0.24 ps, $N_{th}$=6.02 × 10$^{20}$ cm$^{-3}$ |
| Refractive index for unexcited material [52, 53] | $\sqrt{\varepsilon_{un}}$ | $\sqrt{11.67316 + \frac{1}{\lambda_L^2} + \frac{0.004482633}{\lambda_L^2 - 1.108205^2}}$, 22 μm > $\lambda_L$ > 2.5 μm $\sqrt{1 + \frac{10.6684293 \lambda^2}{\lambda_L^2 - 0.301516485^2} + \frac{0.0030434748 \lambda^2}{\lambda_L^2 - 1.13475115^2} + \frac{1.54133408 \lambda^2}{\lambda_L^2 - 1104^2}}$, for 2.5 μm > $\lambda_L$ > 1.36 μm |

**a. Impact of Kerr effect on ultrafast dynamics**

The ultrafast dynamics and the thermal response of the heated material is investigated at three representative laser wavelengths $\lambda_L$ (2.2 μm, 2.5 μm and 3.3 μm) for $\tau_p$=100 fs for (peak fluence) $E_p$ = 100 mJ/cm$^2$. The selection of the wavelengths was based on the excitation properties at these wavelengths; more specifically, $\beta_{TPA} \neq 0$ at 2.2 μm while three-photon absorption dominates the carrier excitation at 2.5 μm and 3.3 μm. Based on the fact that the influence of therefore it does not substantially affect the energy absorption. This behaviour is illustrated both for small (up to the moment when the laser intensity is maximum) and for larger time ranges. It is noted that the reflectivity initially decreases during the pulse duration as the carrier density increases; then, its value ascends rapidly before the end of the pulse followed by a fast decrease and a final slow increase to the initial value (Fig.2a). Similar evolution has been reported for lower wavelengths [25, 36, 47].

The evolution of the carrier density and the carrier and lattice temperatures are illustrated in Fig.3a. The maximum of $T_e$ and $N_e$ occurs shortly after the peak of the pulse. Interestingly, at low intensities (near the left tail of the Gaussian pulse), the carrier temperature does not exhibit a similar behaviour to that demonstrated for Silicon or other semiconductors (i.e. 'a clamped region' which is represented by an initial increase followed by firstly a slight rise and then, a sharp increase) [16, 25, 36, 49, 50]. This is due to the fact that the initial rise is attributed firstly to a single photon absorption (which is absent for laser beam frequencies in the mid-IR range) and secondly to a free carrier absorption which for low intensities is zero for excitation with mid-IR pulses at small timepoints. In regard to the latter, Fig.3b illustrates that $\alpha_{FCA}$ gradually increases to nonzero values during the pulse duration yielding a penetration depth which drops to ~62.5 nm when the laser is turned off. The rise of $\alpha_{FCA}$ due to the increase of the excited carrier density is expected to influence the amount of the absorbed energy. On the other hand, the relative large free carrier absorption at the end of the pulse combined with the large absorption depth through direct (two- and three-photon) absorption mechanisms suggest there is not sufficient time for the carrier and heat transport terms to change substantially the carrier distribution that has been produced. Hence, neglecting the diffusion terms in Eqs.1-2 is not expected to yield different overall behaviour of the thermal response of the irradiated material.

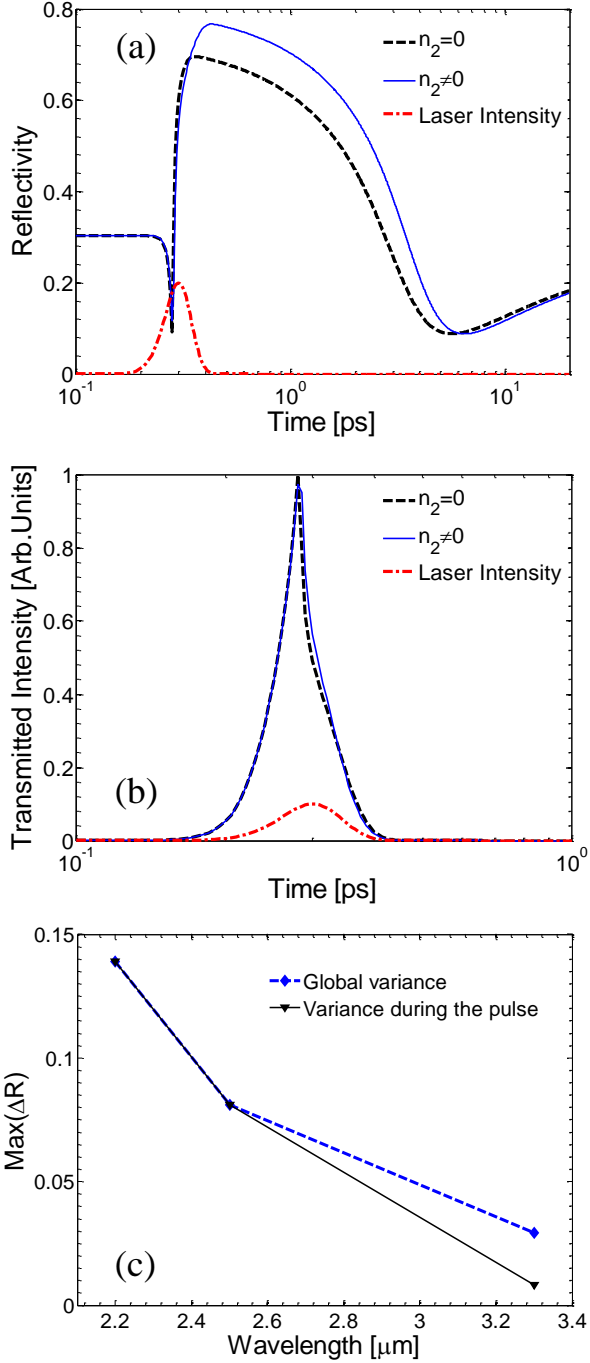

FIG. 2 (Color Online): (a) Dependence of reflectivity on Kerr effect. The laser intensity is sketched in arbitrary units. ($\lambda_L = 2.2$ μm). (b) Effect of Kerr effect on the intensity transmitted through the surface ($\lambda_L = 2.2$ μm) (normalized to 1). (c) Maximum change of reflectivity (for $n_2=0$ and $n_2 \neq 0$) as a function of the wavelength. The *solid* line correspond to variance of max($\Delta R$) in the whole range of timepoints while the *dashed* line represent max($\Delta R$) up to the timepoint where the laser intensity is maximum ($E_p = 0.1$ J/cm$^2$ and $\tau_p = 100$ fs, at $x=y=z=0$).

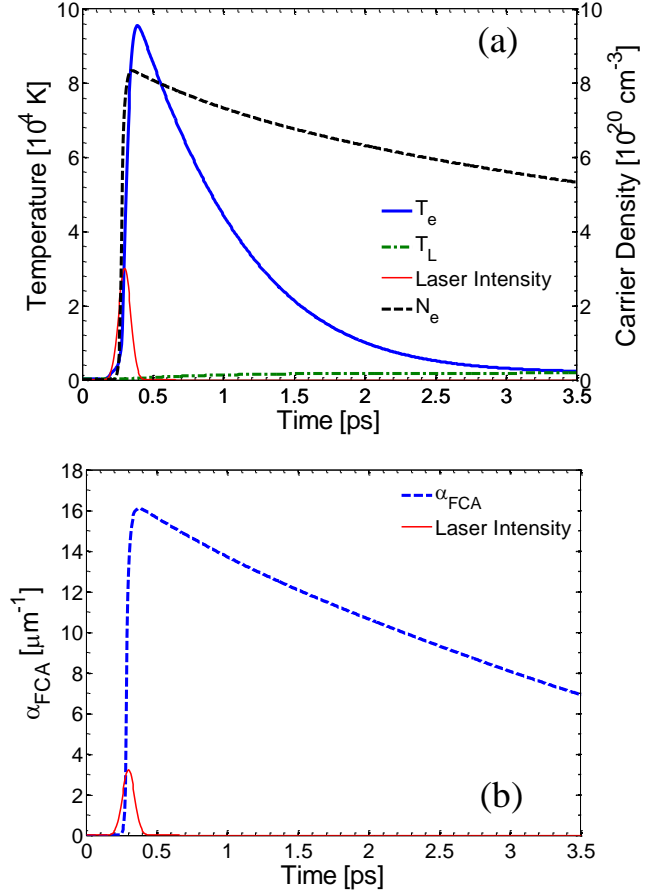

FIG. 3 (Color Online): (a) Evolution of $T_e$, $T_L$, $N_e$ for $n_2 \neq 0$ (at $x=y=z=0$). (b) Absorption coefficient evolution. The laser intensity in both graphs is sketched in arbitrary units. ($E_p = 0.1$ J/cm$^2$, $\tau_p = 100$ fs, $\lambda_L = 2.2$ μm).

Simulations results show that Kerr effect leads always to larger (maximum) values of $N_e$ (Fig.4a). It is evident that due to the variable absorption levels at different nonlinearities. It is noted that while the carrier dynamics analysis for smaller wavelengths ($\lambda_L = 2.2$ μm) yields a maximum $N_e$ larger by about 17% than the value obtained if Kerr effect is ignored, the discrepancy is rather insignificant for larger wavelengths ($\lambda_L = 3.3$ μm). The decrease of the maximum value of the carrier density with increasing $\lambda_L$ is due to the impact of the ionization processes: at $\lambda_L = 2.2$ μm, there is a dominant two-photon absorption assisted ionization while at larger wavelengths, $\beta_{TPA} = 0$ (for $\lambda_L = 2.5$ μm) which enhances the impact of the three photon absorption. At even larger wavelengths ($\lambda_L = 3.3$ μm), the main ionization mechanism, the three-photon assisted ionization process becomes very small due to the decrease of $\gamma_{TPA}$ as $\lambda_L$ increases [58].

Similar discrepancies to the one shown above as a result of the Kerr effect occur also for the electron and lattice wavelengths, the induced density of the excited carriers is larger if the nonlinearities due to Kerr effect are included in the model (Fig.4a). As expected, at larger wavelengths, the Kerr effect impact on the carrier density evolution is insignificant (Fig.4b) due to the negligent influence of Kerr temperatures. Fig.5a shows the rise of the lattice temperature that is demonstrated for $n_2\neq0$. It is evident that the Kerr effect causes a significant thermal response of the material, which is reflected by the enhanced lattice temperature (~200 K for $E_p = 100$ mJ/cm$^2$ at $\lambda_L = 2.2$ μm while similar conclusions can also be drawn for different conditions). This behaviour is of paramount importance as it is expected to influence the damage threshold values (see Section IV). To evaluate the impact of the laser wavelength, it turns out that the Kerr effect does not induce any significant change of the maximum lattice temperature at longer wavelengths (Fig.5b). Furthermore, the effect on the equilibration time or the evolution of $T_e$ (Fig.5a) appears to be insignificant.

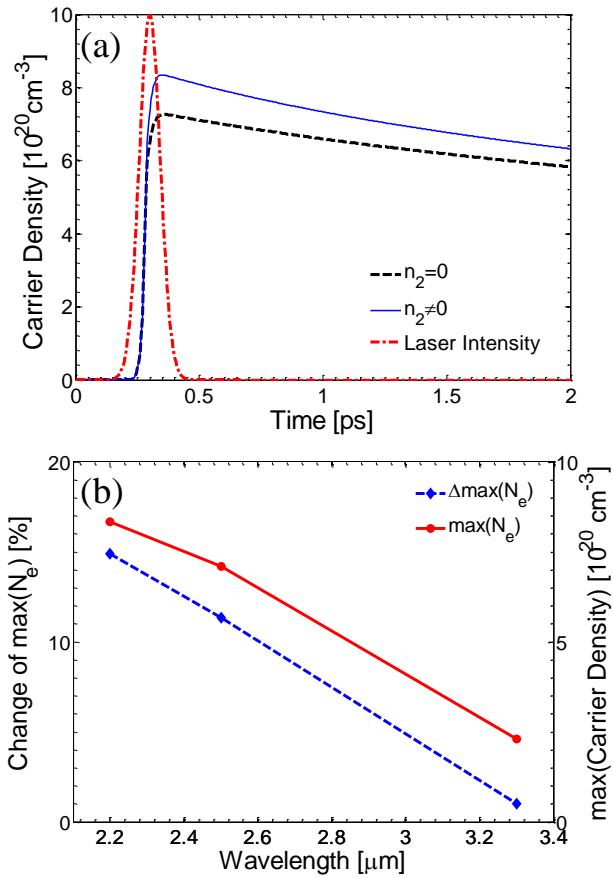

FIG. 4 (Color Online): (a) Dependence of evolution of $N_e$ on Kerr effect. The laser intensity is sketched in arbitrary units ($\lambda_L = 2.2$ μm). (b) Percentage of change of max($N_e$) (for $n_2=0$ and $n_2\neq0$) as a function of the wavelength. The maximum $N_e$ is also illustrated. ($E_p = 0.1$ J/cm$^2$ and $\tau_p = 100$ fs at $x=y=z=0$).

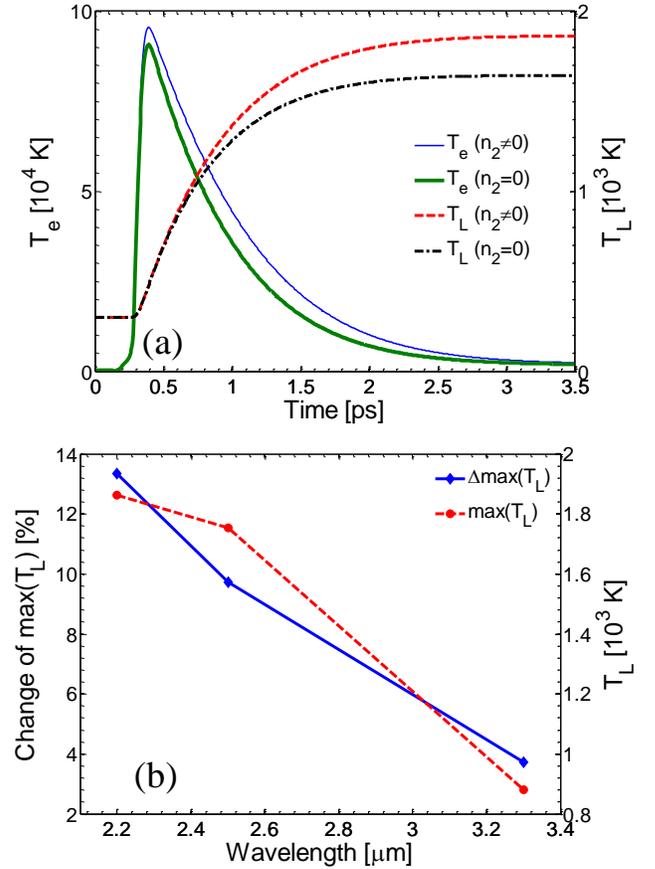

FIG. 5 (Color Online): (a) Dependence of evolution $T_e$ and $T_L$ on Kerr effect. (b) Percentage change of max($T_L$) (for $n_2=0$ and $n_2\neq0$) as a function of the wavelength. ($E_p = 0.1$ J/cm$^2$, $\lambda_L = 2.2$ μm and $\tau_p = 100$ fs at $x=y=z=0$).

### b. Surface plasmon excitation

The condition $Re(\varepsilon) < -1$ and solution of Eqs.1-9 yield the value of the SP wavelength as a function of the carrier densities and the fluence (Fig.6) for $\tau_p = 100$ fs for $\lambda_L = 2.2$ μm, 2.5 μm, and 3.3 μm. According to the theoretical predictions (Fig.6a), a larger (maximum value of) carrier density is required to initiate SP excitation (at $Re(\varepsilon) < -1$) as the laser wavelength decreases. On the other hand, an increase of the laser wavelength leads to smaller overall expected variation of the SP periodicity. More specifically, a comparison with results for larger laser frequencies ($\lambda_L$ = 800 nm) manifests that for $\lambda_L$ = 2.2 μm, 2.5 μm, 3.3 μm, an average maximum drop of the $\lambda_S$ yields values ~0.91 $\lambda_L$ which is higher than the estimate for $\lambda_L$ = 800 nm (~0.78 $\lambda_L$). Based on the results illustrated in Fig.6, a large deviation of the ripple frequency is not expected for irradiation in the mid-IR spectral region unlike predictions for $\lambda_L$ = 800 nm. Furthermore, given the impact of the absorbed energy on the production of excited carriers, the SP periodicity was also calculated as a function of the fluence (Fig.6b). While at $\lambda_L$ = 3.3 μm, a larger peak fluence is necessary to excite SP, for $\lambda_L$ = 2.5 μm, a smaller fluence is required compared to that for $\lambda_L$ = 2.2 μm.

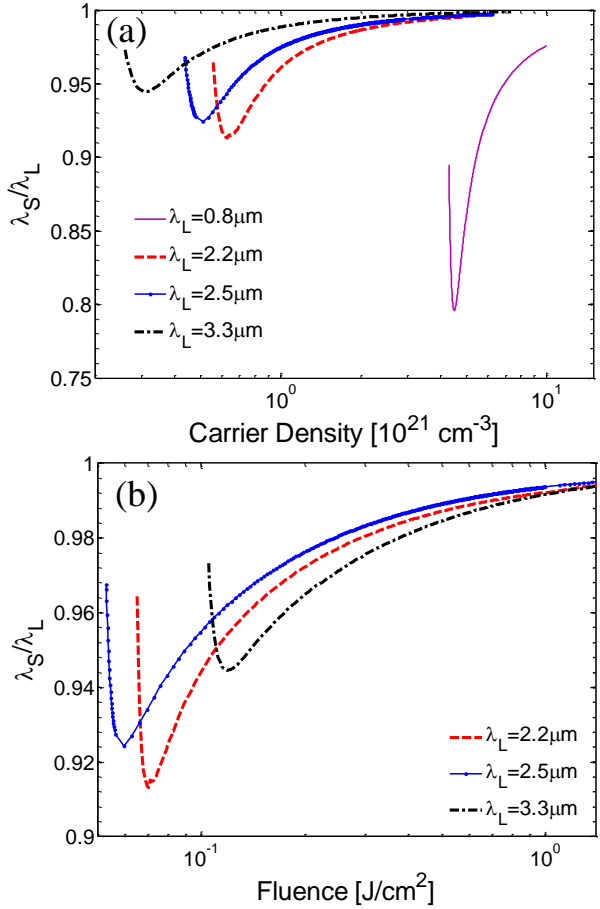

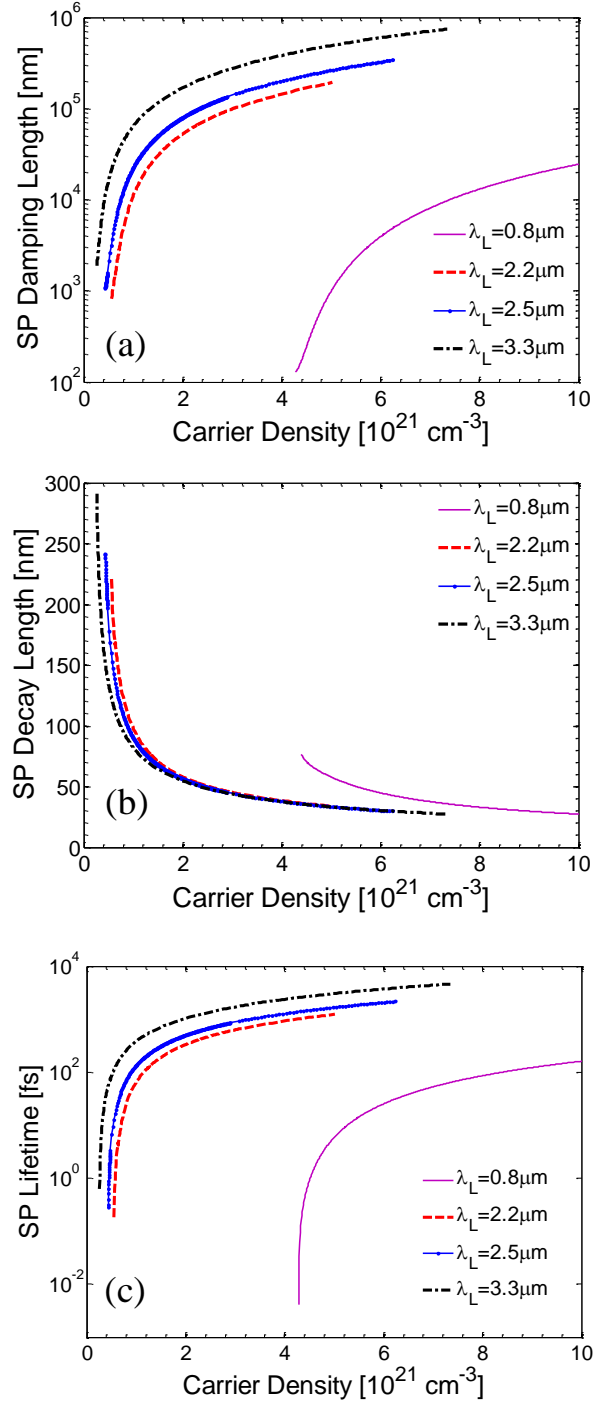

FIG. 6 (Color Online): Surface plasmon periodicity as a function of the (a) carrier density, and (b) fluence at different wavelengths. ($\tau_p = 100$ fs).

FIG. 7 (Color Online): Surface plasmon's (a) damping length, (b) decay length, and (c) lifetime at different laser wavelengths. ($\tau_p = 100$ fs).

As excitation of surface plasmons and their interference with the incident beam is regarded as the predominant mechanism of periodic structure formation [16, 37], the range of the computed SP wavelengths can provide an estimate of the expected sub-wavelength structures on the irradiated material for different $N_e$. The above results highlight significant dfferences of the SP periodicities produced with mid-IR and lower wavelengths which is expected to affect LIPSS modulation.

Apart from the SP wavelength, it is also important to compare spatial features of these surface waves such as their damping length for their propagation along the surface, decay of the SP away from the surface as well as their lifetime. More specifically:

(i) the range of values of the distance which the SP propagates along the surface is computed by the expression $L = \left[ 2Im\left( \sqrt{\frac{\varepsilon \varepsilon_d}{\varepsilon + \varepsilon_d}} \right) \right]^{-1}$ [45]. While for $\lambda_L$ = 800 nm, $L$ ranges from 0.130 μm to 25 μm, substantially larger values (one order of magnitude) between 0.830 μm and 193 μm, 1 μm and 340 μm, and 2 μm and 740 μm are predicted for $\lambda_L$ = 2.2 μm, 2.5 μm, and 3.3 μm, respectively (Fig.7a).

(ii) On the other hand, the decay of the SP away from the surface of the material is given by

$$L_D = \frac{\lambda_L}{2\pi Im\left( \sqrt{\frac{\varepsilon^2}{\varepsilon_d + \varepsilon}} \right)} \tag{10}$$

for the 1/e decay of the electric field based assuming continuity of the electromagnetic field on the surface [57]. Results illustrated in Fig.7b indicate a weak confinement of SP for wavelengths in the mid-IR region compared to the calculations for $\lambda_L$ = 800 nm. It is evident that for carrier densities small enough but sufficiently high to initiate SP excitation, the decay length of the SP electric field inside the material is larger than 220 nm which is three times the estimate for excitation with $\lambda_L$ = 800 nm. The increasing monotonicity of the $L$ at larger wavelengths manifests that the confinement can be further weakened by using even smaller laser wavelengths. By contrast, it is noted that at higher excitation levels in which larger carrier densities are produced, $L_D$ values are comparable regardless of the light frequency used.

(iii) Finally, the lifetime of the SP is calculated by the expression [60]

$$\tau_{SP} = \frac{1}{2A} \tag{11}$$

$$A = \frac{c}{2} \times \frac{2\pi}{\lambda_L} \frac{Re\sqrt{\left( \frac{\varepsilon \varepsilon_d}{\varepsilon + \varepsilon_d} \right)} Im(\varepsilon)}{Re(\varepsilon)} \times \frac{Re(\varepsilon_d)}{Re(\varepsilon) + Re(\varepsilon_d)}$$

Results in Fig.7c illustrate that there is an increase in the SP lifetime if mid-IR frequencies are used compared to predictions for $\lambda_L$ = 800 nm. Interestingly, $\tau_{SP}$ is between one to two orders of magnitude larger for $\lambda_L$ = 2.2 μm, 2.5 μm, and 3.3 μm. More specifically, $\tau_{SP}$ range extends to some picoseconds for mid-IR unlike for $\lambda_L$ = 800 nm for which the SP lifetime lasts up to some hundreds of femtoseconds at large $N_e$. A similar increase in the SP lifetime for longer wavelengths has been reported in previous works for metals [60]. Given the significance of the lifetime of SP for the interference of the incident beam with surface plasmons [16, 27, 36], an increase of $\tau_{SP}$ appears to allow longer pulse temporal separation in double pulse laser-assisting technique to modify the surface profile of a material [33].

c. **Damage threshold**

One important parameter, both from fundamental and applied point of view is the determination of the material damage threshold, $E_{DT}$. In principle, there is an ambiguity of the definition of the damage threshold on whether mass removal is involved or, simply, a mass redistribution (i.e. due to melting and fluid transport) occurs. In the current work, the investigation focuses on the laser conditions that produce effects that raise the temperature of the lattice above the melting point but material volume does not vary. Therefore, $E_{DT}$ is defined as the minimum fluence required to melt the material ($T_L > T_{melt}$).

The dependence of $E_{DT}$ as a function of the pulse duration and wavelength is illustrated in Fig.8a. It is noted that a pulse duration increase leads to a decrease of the absorbed energy which is reflected also on the reduced number of excitation carriers shown in the Supplementary Material [61]. This behaviour indicates that more energy is required to melt the material (i.e. $E_{DT}$ increases). Simulation results show that the damage threshold varies as $\sim \tau_p^{\zeta(\lambda_L)}$ for pulse durations in the range [0.2 ps, 10 ps], where $\zeta(\lambda_L) \sim 0.552$, 0.562, 0.553 for $\lambda_L = 2.2$ μm, 2.5 μm, 3.3 μm, respectively). Similar power law dependencies have been deduced for Silicon and other materials after irradiation at lower wavelengths [49, 62-65]. On the other hand, it is noted that (inset in Fig.8a) for $\tau_p < 200$ fs, there is a deviation from the aforementioned power law dependency due to the influence of other characteristic times in the heating process such as the recombination and relaxation times. This behaviour is more enhanced at lower wavelengths $\lambda_L = 2.2$ μm and $\lambda_L = 2.5$ μm.

To elaborate further on the dependence of $E_{DT}$ on the laser wavelength, it is important to examine this correlation with respect to the strength of the multiphoton excitation mechanisms as well as the factors that are capable to increase carrier excitation (i.e. impact ionisation). More specifically, two distinct cases were analysed: (i) $\beta_{TPA} = 0.25$ cm/GW, $\gamma_{TPA} = 0.020$ cm$^3$/GW$^2$ (for $\lambda_L = 2.2$ μm), (ii) $\beta_{TPA} = 0$ cm/GW, $\gamma_{TPA} = 0.0276$ cm$^3$/GW$^2$ (for $\lambda_L = 2.5$ μm) and (iii) $\beta_{TPA} = 0$ cm/GW, $\gamma_{TPA} = 0.0017$ cm$^3$/GW$^2$ (for $\lambda_L = 3.3$ μm) (Table I and [58]). Interestingly, previous reports in Silicon showed that for near infrared pulses, avalanche (impact) ionisation processes is the driving force for surface damage which leads to lower damage threshold for longer wavelengths [25]; that behaviour could not be explained if one-photon absorption mechanisms accounted for the damage [66]. To investigate whether similar arguments hold for irradiation with pulses in the mid-IR range, simulations have been also carried out [61]. Results provide a compelling proof that impact ionisation itself is sufficient to explain the initially surprising results that radiation with $\lambda_L = 2.5$ μm requires less fluence to damage the material than laser sources of $\lambda_L = 2.2$ μm; by contrast, for irradiation with beams of $\lambda_L = 3.3$ μm in which a three photon absorption mechanism dominates, melting of the material occurs at higher fluences. The effect of impact ionization contribution to the (maximum) carrier change rate (quantity $\theta N_e$ in Eq.1) is illustrated as a function of the laser wavelength and the pulse duration for the damage threshold fluences. Thus, simulation results manifest in a conclusive way that the level of carrier density excitation due to, predominantly, impact ionization produces different behaviour for the three wavelengths at higher $\tau_p$; this justifies the aforementioned argument that avalanche effects do contribute to higher damage thresholds for $\lambda_L = 2.2$ μm than for $\lambda_L = 2.5$ μm. To explain, further though, the damage threshold curves for $\lambda_L = 2.2$ μm and $\lambda_L = 2.5$ μm, one has also to look at the absorption rates for the two- and three-photon absorption mechanisms as well as the strength of $\gamma_{TPA}$. To estimate the contribution of the two- and three-photon absorption in the produced excited carrier distribution, simulations have been performed for $E_p = 0.1$ J/cm$^2$ for six different pulse durations (in the range $\tau_p=0.1$ ps to 0.6ps) for $\lambda_L = 2.2$ μm and $\lambda_L = 2.5$ μm. Results in Fig.8b illustrate the maximum $T_L$ where simulations for the complete model ($\beta_{TPA} = 0.25$ cm/GW, $\gamma_{TPA} = 0.020$ cm$^3$/GW$^2$) are shown for $\lambda_L = 2.2$ μm and it is compared with $\beta_{TPA} = 0$ cm/GW, $\gamma_{TPA} = 0.020$ cm$^3$/GW$^2$. All cases are tested against theoretical results for $\lambda_L = 2.5$ μm ($\beta_{TPA} = 0$ cm/GW, $\gamma_{TPA} = 0.0276$ cm$^3$/GW$^2$). It is evident that for irradiation with for $\lambda_L = 2.2$ μm, the two-photon absorption does not influence the maximum lattice temperature for $\tau_p$> 300 fs (for $E_p$=0.1 J/cm$^2$) and three-photon absorption dominates the excitation from the valence to the conduction band; by contrast it appears that the significance of the two-photon absorption rate is more enhanced at decreasing pulse duration. Similar conclusions can be reached if higher intensities can be achieved by increasing the fluence. More specifically, at higher fluences, two-photon processes play a more important role which is also projected on the produced enhanced lattice temperature with respect to the one for $\beta_{TPA}$ = 0 cm/GW, $\gamma_{TPA}$ = 0.0276 cm$^3$/GW$^2$ for $\lambda_L$ = 2.5 μm. On the other hand, as the pulse duration increases and exceeds $\tau_p$ = 300 fs, the (almost minimum) influence of the two-photon absorption processes yield excitation predominantly through a three-photon absorption with $\gamma_{TPA}$ = 0.020 cm$^3$/GW$^2$; as this value gives smaller three-photon absorption rate for $\lambda_L$ = 2.5 μm, the produced $T_L$ is expected to be smaller than that for for $\lambda_L$ = 2.5 μm. This argument in conjunction with the enhanced impact ionization for $\lambda_L$ = 2.5 μm can be used to explain the higher maximum lattice temperatures for $\lambda_L$ = 2.5 μm than for $\lambda_L$ = 2.2 μm which in turn accounts for the predicted larger threshold for $\lambda_L$ = 2.2 μm (Fig.8a). By contrast, the above argument does not vary the order of the magnitudes of $\lambda_L$ = 3.3 μm and $\lambda_L$ = 2.5 μm (Fig.8a) as the three-photon absorption process for the latter wavelength is always stronger. This is also reflected on the impact ionization contribution to the excited carrier densities [61].

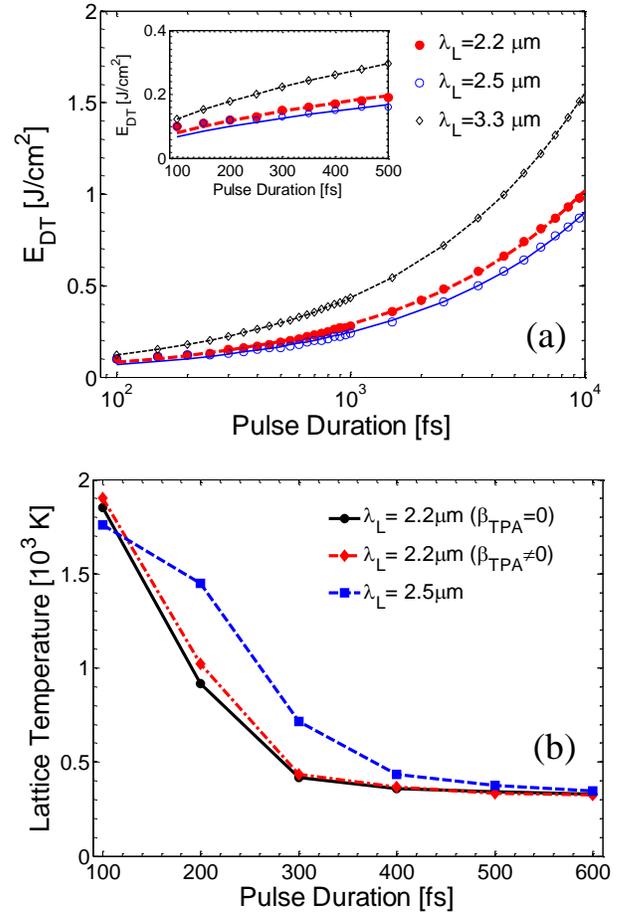

FIG. 8 (Color Online): (a) Damage threshold fluences $E_{DT}$ vs pulse duration. The inset shows an enhanced view of $E_{DT}$ for $\tau_p$ in the range [100 fs, 500 fs] (*points* represent the simulated data values while *lines* are derived after fitting using a power law $\tau_p^{\zeta(\lambda_L)}$, i.e. *thick dashed* line for $\lambda_L = 2.2$ μm, *solid* line for $\lambda_L = 2.5$ μm, and *thin dashed* line for $\lambda_L = 3.3$ μm), (b) Lattice temperature for two- and three-photon absorption for $\lambda_L = 2.2$ μm and $\lambda_L = 2.5$ μm at $E_p=0.1$ J/cm$^2$. ($x=y=z=0$).

Certainly, it has to be noted that the aforementioned predictions require validation of the model with experimental results. Some experimental observations verify the agreement with our predictions for small wavelengths in the mid-IR region (for $\lambda_L$ = 2.2 μm [67]); nevertheless, a more accurate conclusion will be drawn if more appropriately developed experimental (for example, time-resolved experimental) protocols are introduced to evaluate the damage thresholds at the onset of the phase transition. To the best of our knowledge, there are not similar reports with experimental results for the frequency range explored in this study.

There are also some yet unexplored issues that need to be addressed (i.e excitation in very short pulses, structural effects in extreme conditions, more accurate behaviour in ablation conditions, formation of voids inside the material after repetitive irradiation, role of incubation effects, etc.) before a complete picture of the physical processes that characterise heating of Silicon with femtosecond mid-IR laser pulses is attained. Nevertheless, the methodology presented in this work aimed to provide a first insight on the fundamental mechanisms in a previously unexplored area. Apart from the importance of elucidating the underlying mechanisms from a physical point of view, a deeper understanding of the thermal response of the material as well as the characteristics of electrodynamic effects (i.e. lifetime and extinction length of SP), will allow a systematic novel surface engineering with strong mid-IR fields.

## V. CONCLUSIONS

A detailed theoretical framework was presented that describes both the ultrafast dynamics and thermal response following irradiation of Silicon with ultrashort pulsed lasers in the mid-IR range. Results demonstrated that the Kerr effect is important at lower wavelengths (~2.2 μm) and it leads to substantially large deviations to the maximum lattice temperature reached that it affects the damage threshold. Furthermore, it is shown that although the heated material is initially transparent, during the duration of the pulse the energy is confined in a less than ~ 100 nm depth.

A systematic analysis of the SP dispersion relation for mid-IR and comparison with results upon excitation with $\lambda_L$=800 nm revealed that irradiation in the mid-IR region yielded SP that are weakly confined on the surface, exhibit longer lifetimes, and propagate on larger areas. These features can be potentially exploited to promote mid-IR-based technology to produce sensors, detectors or to present new capabilities in laser-based manufacturing.

Finally, theoretical predictions, also, revealed a $\tau_p^{\zeta(\lambda_L)}$ ($\zeta(\lambda_L)$~0.55) dependence of the damage threshold for $\tau_p$> 100 fs. Moreover, analysis for $\lambda_L$=2.2 μm, manifests conclusively the enhanced role of the impact ionization contribution at longer pulse durations which yield eventually to a lower damage threshold for irradiation with laser pulses of $\lambda_L$=2.5 μm. Predictions resulting from the above theoretical approach demonstrate that unravelling new phenomena in the interaction of matter with mid-IR pulses can potentially set the basis for the development of new tools for non-linear optics and photonics for a large range of applications.

## ACKNOWLEDGEMENTS

The authors acknowledge financial support from *Nanoscience Foundries and Fine Analysis (NFFA)–Europe* H2020-INFRAIA-2014-2015 (under Grant agreement No 654360) and *MouldTex* project-H2020-EU.2.1.5.1 (under Grant agreement No 768705).

*Corresponding authors: tsibidis@iesl.forth.gr

[1] A. Y. Vorobyev and C. Guo, Laser & Photonics Reviews **7**, 385 (2012).
[2] V. Zorba, L. Persano, D. Pisignano, A. Athanassiou, E. Stratakis, R. Cingolani, P. Tzanetakis, and C. Fotakis, Nanotechnology **17**, 3234 (2006).
[3] V. Zorba, E. Stratakis, M. Barberoglou, E. Spanakis, P. Tzanetakis, S. H. Anastasiadis, and C. Fotakis, Advanced Materials **20**, 4049 (2008).
[4] D. Bäuerle, *Laser processing and chemistry* (Springer, Berlin; New York, 2000), 3rd rev. enlarged edn.
[5] J.-C. Diels and W. Rudolph, *Ultrashort laser pulse phenomena : fundamentals, techniques, and applications on a femtosecond time scale* (Elsevier / Academic Press, Amsterdam ; Boston, 2006), 2nd edn.
[6] E. L. Papadopoulou, A. Samara, M. Barberoglou, A. Manousaki, S. N. Pagakis, E. Anastasiadou, C. Fotakis, and E. Stratakis, Tissue Eng Part C-Me **16**, 497 (2010).
[7] Z. B. Wang, M. H. Hong, Y. F. Lu, D. J. Wu, B. Lan, and T. C. Chong, Journal of Applied Physics **93**, 6375 (2003).
[8] R. Böhme, S. Pissadakis, D. Ruthe, and K. Zimmer, Applied Physics A-Materials Science & Processing **85**, 75 (2006).
[9] S. M. Petrovic, B. Gakovic, D. Perusko, E. Stratakis, I. Bogdanovic-Radovic, M. Cekada, C. Fotakis, and B. Jelenkovic, Journal of Applied Physics **114**, 233108 (2013).
[10] C. Simitzi, P. Efstathopoulos, A. Kourgiantaki, A. Ranella, I. Charalampopoulos, C. Fotakis, I. Athanassakis, E. Stratakis, and A. Gravanis, Biomaterials **67**, 115 (2015).
[11] E. Stratakis, A. Ranella, and C. Fotakis, Biomicrofluidics **5**, 013411 (2011).
[12] J. Bonse, M. Munz, and H. Sturm, Journal of Applied Physics **97**, 013538 (2005).
[13] M. Huang, F. L. Zhao, Y. Cheng, N. S. Xu, and Z. Z. Xu, ACS Nano **3**, 4062 (2009).
[14] J. E. Sipe, J. F. Young, J. S. Preston, and H. M. Vandriel, Physical Review B **27**, 1141 (1983).
[15] Z. Guosheng, P. M. Fauchet, and A. E. Siegman, Physical Review B **26**, 5366 (1982).
[16] G. D. Tsibidis, M. Barberoglou, P. A. Loukakos, E. Stratakis, and C. Fotakis, Physical Review B **86**, 115316 (2012).
[17] G. D. Tsibidis, E. Skoulas, A. Papadopoulos, and E. Stratakis, Physical Review B **94**, 081305(R) (2016).
[18] G. D. Tsibidis, E. Stratakis, and K. E. Aifantis, Journal of Applied Physics **112**, 089901 (2012).
[19] A. Papadopoulos, E. Skoulas, G. D. Tsibidis, and E. Stratakis, Applied Physics A **124**, 146 (2018).
[20] G. D. Tsibidis and E. Stratakis, Journal of Applied Physics **121**, 163106 (2017).
[21] G. D. Tsibidis, A. Mimidis, E. Skoulas, S. V. Kirner, J. Krüger, J. Bonse, and E. Stratakis, Applied Physics A **124**, 27 (2017).
[22] M. Birnbaum, Journal of Applied Physics **36**, 3688 (1965).
[23] J. Bonse, S. Höhm, S. V. Kirner, A. Rosenfeld, and J. Krüger, Ieee J Sel Top Quant **23**, 9000615 (2017).
[24] Y. Shimotsuma, P. G. Kazansky, J. R. Qiu, and K. Hirao, Physical Review Letters **91**, 247405 (2003).
[25] H. M. Vandriel, Physical Review B **35**, 8166 (1987).
[26] S. K. Sundaram and E. Mazur, Nature Materials **1**, 217 (2002).
[27] T. J. Y. Derrien, J. Kruger, T. E. Itina, S. Hohm, A. Rosenfeld, and J. Bonse, Applied Physics a-Materials Science & Processing **117**, 77 (2014).
[28] E. Knoesel, A. Hotzel, and M. Wolf, Physical Review B **57**, 12812 (1998).
[29] D. S. Ivanov and L. V. Zhigilei, Physical Review B **68**, 064114 (2003).
[30] Z. Lin, L. V. Zhigilei, and V. Celli, Physical Review B **77**, 075133 (2008).
[31] A. Rudenko, J.-P. Colombier, and T. E. Itina, Physical Review B **93**, 075427 (2016).
[32] J. Bonse, A. Rosenfeld, and J. Krüger, Journal of Applied Physics **106**, 104910 (2009).
[33] M. Barberoglou, G. D. Tsibidis, D. Gray, E. Magoulakis, C. Fotakis, E. Stratakis, and P. A. Loukakos, Applied Physics A: Materials Science and Processing **113**, 273 (2013).
[34] G. D. Tsibidis, E. Stratakis, P. A. Loukakos, and C. Fotakis, Applied Physics A **114**, 57 (2014).
[35] J. JJ Nivas, S. He, A. Rubano, A. Vecchione, D. Paparo, L. Marrucci, R. Bruzzese, and S. Amoruso, Sci Rep-Uk **5**, 17929 (2015).
[36] A. Margiolakis, G. D. Tsibidis, K. M. Dani, and G. P. Tsironis, Physical Review B **98**, 224103 (2018).

[37]	T. J. Y. Derrien, T. E. Itina, R. Torres, T. Sarnet, and M. Sentis, Journal of Applied Physics **114**, 083104 (2013).
[38]	O. Varlamova, F. Costache, J. Reif, and M. Bestehorn, Applied Surface Science **252**, 4702 (2006).
[39]	D. R. Austin *et al.*, Journal of Applied Physics **120**, 143103 (2016).
[40]	D. R. Austin *et al.*, Optics Express **23**, 19522 (2015).
[41]	R. Stanley, Nat Photonics **6**, 409 (2012).
[42]	A. D. Bristow, N. Rotenberg, and H. M. van Driel, Applied Physics Letters **90**, 191104 (2007).
[43]	F. Gholami, S. Zlatanovic, A. Simic, L. Liu, D. Borlaug, N. Alic, M. P. Nezhad, Y. Fainman, and S. Radic, Applied Physics Letters **99**, 081102 (2011).
[44]	T. J. Y. Derrien, J. Krüger, and J. Bonse, Journal of Optics **18**, 115007 (2016).
[45]	H. Raether, *Surface plasmons on smooth and rough surfaces and on gratings* (Springer-Verlag, Berlin ; New York, 1988), Springer Tracts in Modern Physics, 111.
[46]	N. K. Hon, R. Soref, and B. Jalali, Journal of Applied Physics **110**, 011301 (2011).
[47]	K. Sokolowski-Tinten and D. von der Linde, Physical Review B **61**, 2643 (2000).
[48]	L. Jiang and H. L. Tsai, Journal of Applied Physics **100**, 023116 (2006).
[49]	J. K. Chen, D. Y. Tzou, and J. E. Beraun, International Journal of Heat and Mass Transfer **48**, 501 (2005).
[50]	A. Rämer, O. Osmani, and B. Rethfeld, Journal of Applied Physics **116**, 053508 (2014).
[51]	S. I. Anisimov, Kapeliov.Bl, and T. L. Perelman, Zhurnal Eksperimentalnoi Teor. Fiz. **66**, 776 (1974 [Sov. Phys. Tech. Phys. 11, 945 (1967)]).
[52]	D. Chandler-Horowitz and P. M. Amirtharaj, Journal of Applied Physics **97**, 123526 (2005).
[53]	C. D. Salzberg and J. J. Villa, Journal of the Optical Society of America **47**, 244 (1957).
[54]	Q. Lin, J. Zhang, G. Piredda, R. W. Boyd, P. M. Fauchet, and G. P. Agrawal, Applied Physics Letters **91**, 021111 (2007).
[55]	D. Dufft, A. Rosenfeld, S. K. Das, R. Grunwald, and J. Bonse, Journal of Applied Physics **105**, 034908 (2009).
[56]	R. M. Osgood, N. C. Panoiu, J. I. Dadap, X. P. Liu, X. G. Chen, I. W. Hsieh, E. Dulkeith, W. M. J. Green, and Y. A. Vlasov, Adv Opt Photonics **1**, 162 (2009).
[57]	J. M. Pitarke, V. M. Silkin, E. V. Chulkov, and P. M. Echenique, Reports on Progress in Physics **70**, 1 (2007).
[58]	S. Pearl, N. Rotenberg, and H. M. van Driel, Applied Physics Letters **93**, 131102 (2008).
[59]	D. Agassi, Journal of Applied Physics **55**, 4376 (1984).
[60]	J. Y. D. Thibault, K. Jörg, and B. Jörn, Journal of Optics **18**, 115007 (2016).
[61]	See Supplementary Material at [URL] for a detailed description of (a) maximum carrier dependence on the pulse duration, (b) temporal variation of the impact ionisation rate, and (c) carrier density as a function of the laser fluence.
[62]	C. S. R. Nathala, A. Ajami, W. Husinsky, B. Farooq, S. I. Kudryashov, A. Daskalova, I. Bliznakova, and A. Assion, Applied Physics A **122**, 107 (2016).
[63]	J. R. Meyer, M. R. Kruer, and F. J. Bartoli, Journal of Applied Physics **51**, 5513 (1980).
[64]	D. P. Korfiatis, K. A. T. Thoma, and J. C. Vardaxoglou, Journal of Physics D-Applied Physics **40**, 6803 (2007).
[65]	J. Bonse, S. Baudach, J. Kruger, W. Kautek, and M. Lenzner, Applied Physics a-Materials Science & Processing **74**, 19 (2002).
[66]	P. P. Pronko, P. A. VanRompay, C. Horvath, F. Loesel, T. Juhasz, X. Liu, and G. Mourou, Physical Review B **58**, 2387 (1998).
[67]	K. Soong, R. L. Byer, C. McGuiness, E. Peralta, and E. Colby, Proceedings of the 2011 Particle Accelerator Conference, (IEEE, New York, 2011), 277.

*Supplementary Material for*

# Ultrafast dynamics and surface plasmon properties of Silicon upon irradiation with mid-infrared femtosecond laser pulses

E. Petrakakis [1,2], G. D. Tsibidis [1*] and E. Stratakis [1,2]

[1] *Institute of Electronic Structure and Laser (IESL), Foundation for Research and Technology (FORTH), N. Plastira 100, Vassilika Vouton, 70013, Heraklion, Crete, Greece*

[2] *Materials Science and Technology Department, University of Crete, 71003 Heraklion, Greece*

### a. Maximum carrier density vs. Pulse duration

The curves in Fig.S1 illustrate the correlation of the (maximum) carrier density as a function of the laser pulse duration at various wavelengths in the mid-IR spectral region. The fluence used to derive these curves is the damage threshold (Fig.8a).

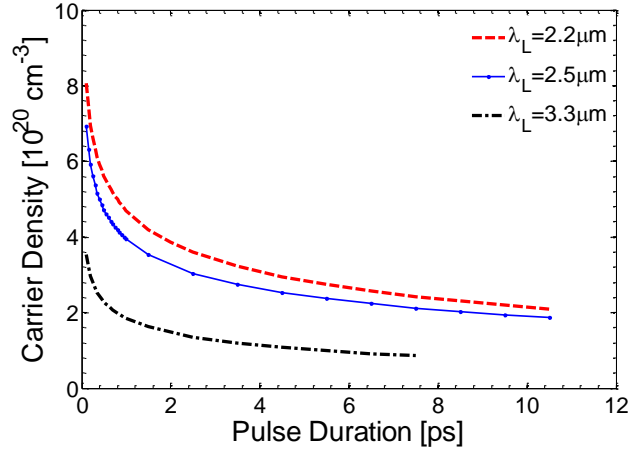

FIG. S1: (Maximum) Carrier density variation as a function of the pulse duration at different wavelengths for damage threshold conditions.

### b. Impact ionization vs. Time

Simulation results for the temporal dependence of the impact ionization ($\theta N_e$) at three different wavelengths and $\tau_p = 100$ fs and $\tau_p = 500$ fs demonstrate the significant contribution to the carrier density changes. More specifically, for $\tau_p = 100$ fs, the impact ionization is larger as the laser wavelength increases. By contrast, for $\tau_p = 500$ fs (and without loss of generality for larger wavelengths), the impact ionization for $\lambda_L=2.2$ μm is smaller than for $\lambda_L=2.5$ μm while for $\lambda_L=3.3$ μm it takes the smallest value. This behaviour presented in Fig.S2 due to the contribution of the impact ionization at different wavelengths is used to explain the larger damage threshold fluence for $\lambda_L$=2.2 μm compared to that for $\lambda_L$=2.5 μm.

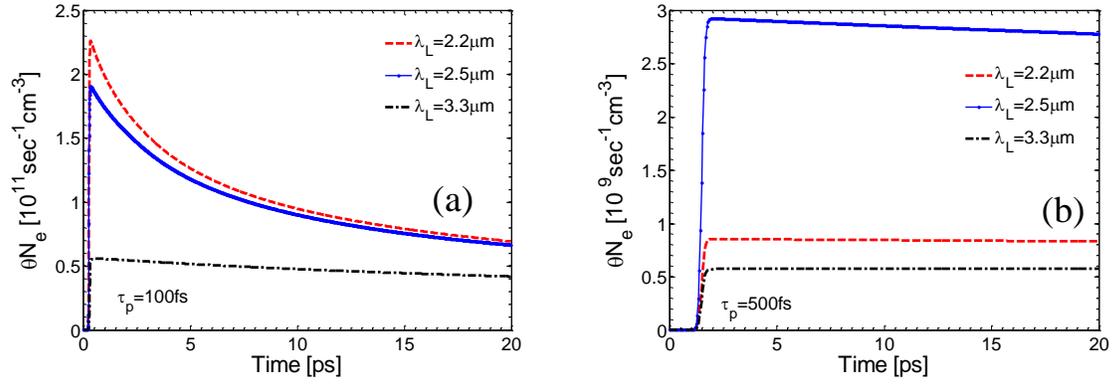

FIG. S2: Impact ionization *vs.* time for $\tau_p$ = 100 fs (a) and $\tau_p$ = 500 fs (a) at different laser wavelengths ($E_p$=0.1 J/cm$^2$).

### c. Carrier density vs. Fluence

The curves in Fig.S3 illustrate the correlation of the carrier density as a function of the laser fluence at various wavelengths in the mid-IR spectral region.

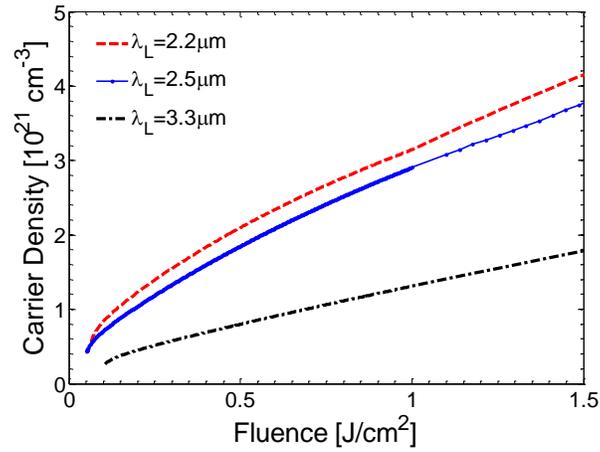

FIG. S3: (Maximum) Carrier density variation as a function of the laser fluence at different wavelengths. ($\tau_p$ = 100 fs).